\documentclass[11pt,notitlepage]{article}
\begin{document}

\title{The direction of time in quantum field theory}
\author{Peter Morgan\\
\emph{Physics Department, Yale University,}\\
\emph{New Haven, CT 06520, USA.}}

\maketitle
\begin{abstract}
The algebra of observables associated with a quantum field theory is invariant under the connected component of
the Lorentz group and under parity reversal, but it is not invariant under time reversal.
If we take general covariance seriously as a long-term goal, the algebra of observables should be
time-reversal invariant, and any breaking of time-reversal symmetry will have to be described by the
state over the algebra.
In consequence, the modified algebra of observables is a presentation of a classical continuous random field.
\end{abstract}

\newcommand\Half{{\raisebox{0.2ex}{$\scriptstyle\frac{\scriptstyle 1}{\raisebox{-0.4ex}{$\scriptstyle 2$}}$}}}

First some mathematical preliminaries are necessary.
Quantum field theory is presented at an elementary level in terms of an operator-valued distribution, $\hat\phi(x)$.
That this is a distribution reflects the fact that $\hat\phi(x)$ is not itself an operator that we can associate
with a measurement; for an operator, we have to \emph{smooth} the quantum field by averaging, to obtain
$\hat\phi_f=\int f(x)\hat\phi(x)d^4x$, where the \emph{test function} $f(x)$ is generally taken to be a Schwartz
space function, which is zero at infinity and smooth both in real space and, as $\tilde f(k)$, in Fourier space.
There are notational, conceptual, and mathematical advantages to working with the \emph{smeared} operators
$\hat\phi_f$ instead of with the operator-valued distribution $\hat\phi(x)$, and we can always get back to
operator-valued distributions, albeit improperly, by using Dirac delta functions.
Routinely, $\hat\phi_f$ is expressed as the sum of non-observable creation and annihilation operators
$a_f^\dagger$ and $a_f$, $\hat\phi_f=a_f+a_{f^*}^\dagger$, where $a_f$ and $a_{f^*}^\dagger$ are both
complex linear in $f$ to ensure that $\hat\phi_f$ is complex linear.
The quantized Klein-Gordon field, for example (because it is the most elementary non-interacting quantum field),
can be straightforwardly presented in terms of commutation relations between creation and annihilation
operators\cite{MorganPhysLettA2},
\begin{equation}
  \Bigl[a_g,a^\dagger_f\Bigr]=(f,g),\qquad
  \Bigl[a_f,a_g\Bigr]=0.
\end{equation}
The Hermitian inner product $(f,g)$ is manifestly Lorentz invariant, except for time reversal,
\begin{equation}\label{InnerProduct}
  (f,g)=\hbar \int\frac{d^4k}{(2\pi)^4}2\pi
         \delta(k^\mu k_\mu-m^2)\theta(k_0)\tilde f^*(k)\tilde g(k).
\end{equation}
This fixes the algebraic structure of the quantized Klein-Gordon field operators.
In particular, this construction ensures that the field $\hat\phi_f$ satisfies microcausality, so that
$[\hat\phi_f,\hat\phi_g]=0$ whenever the real-space supports of the functions $f$ and $g$ are space-like
separated (in terms of operator-valued distributions, $[\hat\phi(x),\hat\phi(y)]=0$ whenever $x$ and $y$ are
space-like separated).
\textbf{The $\theta(k_0)$ factor, however, which implements the requirement that the energy spectrum of the
Hamiltonian operator must be positive, introduces an explicit direction of time into quantum field theory.}
The Hamiltonian operator is the generator of time translations in quantum theory, and is required to be in
the forward time-like direction, so this is not a surprise, but it deserves attention.
Note that because of the connection with time translations and the positivity of the Hamiltonian,
non-invariance under time-reversal transcends the straightforward free quantum field model that is
described above.

The nature of this direction of time in quantum field theory has perhaps been less considered than it might
have been because of the ways in which quantum field theory is usually presented.
Especially curious is what happens if we change the inner product so that there is no explicit direction of time,
\begin{equation}\label{InnerProductNDT}
  (f,g)_C=\Half\hbar \int\frac{d^4k}{(2\pi)^4}2\pi
         \delta(k^\mu k_\mu-m^2)\tilde f^*(k)\tilde g(k)=\Half\left[(f,g)+(g^*,f^*)\right].
\end{equation}
In consequence of this choice, the quantum field $\hat\phi_f$ becomes classical --- in fact a
presentation of a continuous random field --- in the sense that
\begin{equation}
  [\hat\phi_f,\hat\phi_g]=[a_f,a_{g^*}^\dagger]+[a_{f^*}^\dagger,a_g]=
  \Half\left[(g^*,f)+(f^*,g)-(f^*,g)-(g^*,f)\right]=0
\end{equation}
\emph{whatever} functions we use for $f$ and $g$.
In this Hilbert space formalism, in other words, the choice of a direction of time is \emph{the} difference
between classical and quantum fields.
If we take it that the algebra of observables \emph{ought} to be invariant under the whole Lorentz group,
including under the discrete parity and time-reversal symmetries --- indeed, it ought to be diffeomorphism
invariant, but that is for another day --- then any lack of parity and time-reversal symmetry should be
described by the state over the algebra of observables.

To consider the meaning of test functions and the projection to positive frequencies in the inner product,
we will specialize to the vacuum sector, which is constructed by defining a vacuum state
$\left|0\right>$ as the zero eigenstate of all annihilation operators, $a_f\left|0\right>=0$ for all test functions.
The vacuum sector, then, is the Fock space of states constructed by applying creation operators to the vacuum,
$a_g^\dagger\left|0\right>$, $a_{g_1}^\dagger a_{g_2}^\dagger\left|0\right>$, ...
(more abstractly, the Fock space can be constructed by using the GNS construction\cite[Ch. 3]{Haag}).

There are two ways in which this algebra of operators can be used in the vacuum sector.
Most obviously, we could measure $\hat\phi_f$ in the vacuum state; for an ensemble of measurements in the
vacuum state we would obtain a probability distribution with moments
$\left<0\right|\hat\phi_f^{2k}\left|0\right>=\frac{(2k)!}{2^k k!}(f^*,f)^k$,
$\left<0\right|\hat\phi_f^{2k-1}\left|0\right>=0$, which correspond to a normal probability distribution
with mean $0$ and variance $(f^*,f)\,$\footnote{$\hat\phi_f$ is only an observable if $\hat\phi_f^\dagger=\hat\phi_f$,
which requires that $f^*=f$ is real, so that $(f,f)_C=(f,f)$; for this observable vacuum classical
and quantum probabilities coincide.}.
This approach, however, is inappropriate for most real measurement apparatuses, which are tuned to give a zero
response to the vacuum.
A different approach, which is almost always used in some variant in quantum optics\footnote{The quantized
electromagnetic field can be constructed exactly as above\cite{MorganJMP,MenikoffSharp}, except that the
inner product includes the components of bivector test functions $f_{\mu\nu}$ and $g_{\mu\nu}$,
\begin{eqnarray*}\label{EMInnerProduct}
     (f,g)_{EM} &=& \hbar\int\frac{d^4k}{(2\pi)^4}
             2\pi\delta(k_\alpha k^\alpha)\theta(k_0)
             \tilde f_{\mu\beta}^*(k)k^\mu k^\nu\tilde g_{\nu}^{\ \beta}(k).
\end{eqnarray*}
The algebraic structure is thus identical above the level of the inner product, but the geometrical
structure in space-time that is expressed by the inner product is different.},
is to use the projection operator
\begin{equation}
  \hat X_f=\frac{a_f^\dagger\left|0\right>\left<0\right|a_f}{(f,f)},
\end{equation}
very often with an improper pure wave-number test function\footnote{We cannot use the $(f,f)$
normalization constant to construct a true projection operator for a pure wave-number test function.
$(f,f)$ is not defined for a delta function in Fourier space, a pure frequency in a single direction
that is evenly distributed over all of space-time.
Note that the commutator $[\hat X_f,\hat X_g]$ is generally non-zero when the supports of $f$ and $g$ are
space-like separated, so quantum optics formalisms which use this or similar operators are not causally
local in this sense.
Nor is $\hat X_f$ linear in $f$.}.
This kind of measurement asks whether a state \emph{resonates} with the measurement apparatus; for example,
in the vacuum state the moments of the probability distribution are all zero, signifying that we
always observe $0$; in the normalized state $\left|g\right>=\frac{1}{\sqrt{(g,g)}}a_g^\dagger\left|0\right>$ the
moments of the probability distribution are all $p=\frac{1}{(g,g)(f,f)}|(f,g)|^2$, signifying that we observe
$1$ with probability $p$ and $0$ with probability $1-p$.
Sometimes the measurement apparatus will resonate, sometimes it won't, depending on how closely parallel
the test functions $g$ and $f$ are in terms of the inner product that defines the algebraic structure.
Quantum optics has constructed many useful states and measurement operators that are used to model experiments,
which will not be further rehearsed here.

Every construction of an observable that is possible in quantum field theory is also
possible for a classical continuous random field, using the classical inner product
$(f,g)_C=\Half\left[(f,g)+(g*,f*)\right]$ instead of using the quantum inner product $(f,g)$;
superpositions and interference are just as possible for continuous random fields as for quantum fields.
What, then, is the difference between the classical and the quantum inner products?
Firstly, the difference between the quantum and classical inner products,
$(f,g)-(f,g)_C=\Half\left[(f,g)-(g*,f*)\right]$, is zero when the supports of $f$ and $g$ are space-like
separated.
Additionally, there is precisely a factor of two between the quantum and classical inner products if
classical modeling uses only test functions that are restricted to positive frequencies (a choice
that results in the \emph{analytic signal} in classical signal analysis, so we may perhaps use the name
\emph{analytic test function}).
With test functions used in classical models restricted to positive frequency, quantum optics and a
classical continuous random field version of quantum optics are operationally identical, albeit with an
inessential factor of $2\,$\footnote{Measurements constitute a set of constraints on the ans\"atze that are
chosen as models for a given set of experiments. If the constraints are satisfiable by $f$, they are also
satisfiable by a constant multiple of $f$.}.
In effect, the continuous random field exploits more degrees of freedom than the corresponding quantum field
theory, and has the same functional dependence on the common degrees of freedom, so it can accommodate
empirical data at least as well.
Note that it is a commonplace in classical signal analysis that the perfect measurement of signal frequency
is incompatible with the measurement of the signal for only a finite time, so that --- for example, because
signal analysis is a large subject --- the Wigner function is a common tool in classical signal
analysis\cite{Cohen}.
I have discussed the differences, similarities, and relationships between the classical and quantum theories
of measurement and their algebras of observables, from a field theory point of view,
elsewhere\cite{MorganPhysLettA2,MorganVaxjo,MorganJMP,MorganStrawMan}\footnote{Of related interest,
Hobson\cite{HobsonAJP,HobsonPT} has recommended using ideas from quantum field theory when motivating
quantum mechanics at the undergraduate level.
In Hobson's approach, however, fields have a particle aspect that causes discrete events, whereas I prefer
to understand events as the result of resonances of the field with the carefully tuned thermodynamic
properties of experimental apparatus that are not point-like.}.

There is a significant sense in which quantum field theory is overconstrained by the restriction to
positive frequency: there are no known interacting quantum field theories on Minkowski space.
In contrast, with the introduction of negative frequencies a large class of interacting continuous random field
theories can be constructed\cite{MorganJMP}, following an approach that was tried but abandoned for quantum
fields in the 1960s.
Hence, there is both a significant mathematical advantage and a significant conceptual advantage to
using classical continuous random field models consistently in Physics.
Of course there may be other constraints that have not yet been considered that will make continuous
random fields either impossible or impractical, or simply on balance not attractive to Physicists, but note
that Bell inequalities are not more problematic than they are for quantum fields\cite{MorganBellRF,MorganStrawMan}.

If we eliminate the direction of time from the algebra of observables, there will presumably 
be a significant breaking of time invariance in the states we construct, for we know that
it is most often possible to model experiments in Physics using only fields, without having to resort
to their time-reversed anti-fields\footnote{Since ``anti-field'' is not to my knowledge an existing
terminology, please substitute ``anti-particle'' if you cannot yet give up particle language.}.
A continuous random field formalism effectively has no anti-fields because the algebra
of observables is already time-reversal invariant.

The immediate consequences for quantum field theory of enforcing time-reversal invariance of the algebra of
observables are extreme: instead of using quantum field models, we use continuous random field models, and
we can use Lie fields to express non-Gaussian vacuum correlations\cite{MorganJMP}, instead of having to resort
to renormalization.
The Lie field approach that is made available when we require time-reversal invariance of the algebra of
observables results in a reconceptualization of Physics that goes far beyond the Nature of Time.
The Lie field approach, however, is essentially an empiricist intermediary for future theories, because
only correlations are explicitly modeled; causality, which is an essential part of the explanatory and
predictive power of a theory but cannot be directly measured in the quantum mechanical world of discrete
measurement events, is only emergently part of a continuous random field model.

\end{document}